\documentclass[5p,sort&compress]{elsarticle}
\usepackage{graphicx}
\usepackage{amsfonts}
\usepackage{amssymb}
\usepackage{amsmath}
\usepackage{float}
\usepackage{color}
\usepackage{ulem}
\usepackage{bm}
\usepackage{longtable}
\usepackage{array}

\newcommand*{\rttensor}[1]{\boldsymbol{#1}}
\newcommand*{\rvector}[1]{\boldsymbol{#1}}

\begin{document}
\begin{frontmatter}
\title{Designing Disordered Hyperuniform Two-Phase Materials with Novel Physical Properties}

\author[chm]{D. Chen}

\author[chm,phy,prism,appl]{S. Torquato\corref{cor1}}
\ead{torquato@electron.princeton.edu}

\cortext[cor1]{Corresponding author}

\address[chm]{Department of Chemistry, Princeton University, Princeton, New Jersey 08544, USA}
\address[phy]{Department of Physics, Princeton University, Princeton, New Jersey 08544, USA}
\address[prism]{Princeton Institute for the Science and Technology of Materials, Princeton University, Princeton, New Jersey 08544, USA}
\address[appl]{Program in Applied and Computational Mathematics, Princeton University, Princeton, New Jersey 08544, USA}


\begin{abstract}
Heterogeneous materials consisting of different phases are ideally suited to achieve a broad spectrum of desirable bulk physical properties by combining the best features of the constituents through the strategic spatial arrangement of the different phases. Disordered hyperuniform heterogeneous materials are new, exotic amorphous matter that behave like crystals in the manner in which they suppress volume-fraction fluctuations at large length scales, and yet are isotropic with no Bragg peaks. In this paper, we formulate for the first time a Fourier-space numerical construction procedure to design at will a wide class of disordered hyperuniform two-phase materials with prescribed spectral densities, which enables one to tune the degree and length scales at which this suppression occurs. We demonstrate that the anomalous suppression of volume-fraction fluctuations in such two-phase materials endow them with novel and often optimal transport and electromagnetic properties. Specifically, we construct a family of phase-inversion-symmetric materials with variable topological connectedness properties that remarkably achieves a well-known explicit formula for the effective electrical (thermal) conductivity. Moreover, we design disordered stealthy hyperuniform dispersion that possesses nearly optimal effective conductivity while being statistically isotropic. Interestingly, all of our designed materials are transparent to electromagnetic radiation for certain wavelengths, which is a common attribute of all hyperuniform materials. Our constructed materials can be readily realized by 3D printing and lithographic technologies. We expect that our designs will be potentially useful for energy-saving materials, batteries and aerospace applications.

\end{abstract}
\end{frontmatter}

\section{Introduction}
Heterogeneous materials that consist of different phases (or constituent materials) abound in nature and synthetic products, such as composites, polymer blends, porous media, and powders \cite{To02, Br90, Pa16, Gl17}. In many instances, the length scale of the inhomogeneities is much smaller than the macroscopic length scale of the material, and microscopically the material can be viewed as a homogeneous material with macroscopic or \textit{effective} properties \cite{To02, Sa03a, Sa03b, Su96, Ka03}. It has been shown that given the individual phases, the effective properties of the materials are uniquely determined by microstructure of the phases \cite{To02}. Consequently, the discovery of novel guiding principles to arrange the constituents presents a promising path to design and realize materials with a broad spectrum of exotic and desirable properties by combining the best features of the constituents. The concept of disordered hyperuniformity provides guiding design principles for the creation of materials with singular performance characteristics, as we will demonstrate in this work. 

The notion of hyperuniformity was first introduced in the context of many-particle systems over a decade ago \cite{To03}. Hyperuniform many-body systems are those characterized by an anomalous suppression of density fluctuations at long wavelengths relative to those in typical disordered systems such as ideal gases, liquids and structural glasses. All perfect crystals and perfect quasicrystals, and certain special disordered systems are hyperuniform \cite{To03, Za09}. Disordered hyperuniform many-particle systems are exotic amorphous state of matter that lie between crystal and liquid states: they behave like crystals in the way that they suppress density fluctuations at very large length scales, and yet they are statistically isotropic with no Bragg peaks. There is a special type of hyperuniformity called disordered stealthy hyperuniformity, characterized by the absence of scattering within a range of small wavenumbers around the origin in the Fourier space \cite{Uc04, Zh15}.

The exotic structural features of disordered hyperuniform systems appear to endow such systems with novel physical properties. For example, disordered hyperuniform dielectric networks were found to possess complete photonic band gaps comparable in size to photonic crystals \cite{Fl09, Ma13}. Interestingly, such networks are isotropic, i.e., electromagnetic radiation propagates through the networks independent of the direction, which is an advantage over photonic crystals, and thus makes them suitable for applications such as lasers, sensors, waveguides, and optical microcircuits \cite{Ma13}. Moreover, disordered hyperuniform patterns can have optimal color-sensing capabilities, as evidenced by avian photoreceptors \cite{Ji14}. Recently it was revealed that the electronic band gap of amorphous silicon widens as it tends toward a hyperuniform state \cite{He13}. In the context of superconductors, it was shown that hyperuniform pinning site geometries exhibit enhanced pinning \cite{Th16}, which is robust over a wide range of parameters. In addition, there is evidence suggesting that disordered hyperuniform particulate media possess nearly optimal transport properties while maintaining isotropy \cite{Zh16}. 

These tantalizing findings have provided an impetus for researchers to discover and/or synthesize new disordered hyperuniform systems. We now know that disordered hyperuniformity arises in both equilibrium and nonequilibrium systems across space dimensions; e.g., maximally random jammed hard-particle packings \cite{Za11a, Za11b, Za11c, At13}, driven nonequilibrium granular and colloidal systems \cite{He15, We15}, dynamical processes in ultracold atoms \cite{Le14}, geometry of neuronal tracts \cite{Bu15}, immune system receptors \cite{Ma15} and polymer-grafted nanoparticle systems \cite{Ch17}. The reader is referred to Refs. \cite{To16a} and \cite{To16b} for a comprehensive review of disordered hyperuniform systems that have been discovered so far.

Recently the concept of disordered hyperuniformity has been generalized to two-phase heterogeneous materials \cite{Za09, To16a, To16b}. These materials possess suppressed volume-fraction fluctuations at large length scales, and yet are isotropic with no Bragg peaks. This can sometimes offer advantages over periodic structures with high crystallographic symmetries in which the physical properties can have undesirable directional dependence \cite{Fl09, Ma13}. Specifically, the spectral density $\widetilde{\chi}_{_V}({\bf k})$ of such system goes to zero as the wavenumber $k$ goes to zero with a power-law scaling \cite{Za09, Za11a, Za11b, Za11c, Dr15}, i.e., 
\begin{equation}
\label{eq_a1} \widetilde{\chi}_{_V}({\bf k}) \sim |{\bf k}|^\alpha,
\end{equation}
where $\alpha$ is the scaling exponent. Equivalently, the local volume-fraction variance $\sigma_{_{V}}^2(R)$ associated with a $d$-dimensional spherical observation window of radius $R$ possesses the following scaling at large $R$ \cite{Za09, Za11a, Za11b, Za11c, Dr15}:
\begin{equation}
\label{eq_a2} \sigma_{_{V}}^2(R) \sim \left \{
\begin{array}{l@{\hspace{0.3cm}}c}
R^{-(d+1)}, & \alpha > 1,
\\ R^{-(d+1)}\ln R, & \alpha = 1,
\\ R^{-(d+\alpha)}, & 0 < \alpha < 1.
\end{array} \right. (R\rightarrow\infty)
\end{equation}
where $d$ is the dimension. Note that in all three cases $\sigma_{_{V}}^2(R)$ decays more rapidly than the inverse of the window volume, i.e., faster than $R^{-d}$, which is different from typical disordered two-phase materials. 

Our ability to design disordered hyperuniform two-phase materials in a systematic fashion is currently lacking and hence their potential for applications has yet to be explored. In this work, we develop for the first time a Fourier-space based numerical construction procedure to design at will a wide range of disordered hyperuniform two-phase materials by tuning the shape of the spectral density function across phase volume fractions. This is equivalent to tuning the degree and length scales at which there is anomalous suppression of volume-fraction fluctuations in these materials. We note that the Fourier-space setting is the most natural one, since hyperuniformity is defined in \sout{the} Fourier space. This setting is crucial for capturing accurately the long-wavelength, or equivalently, small-wavenumber $k$ behavior. Our designed disordered hyperuniform microstructures include ones with phase-inversion symmetry as well as a stealthy dispersion. We compute the two-point cluster function, which incorporates nontrivial topological connectedness information and is known to provide a discriminating signature of different microstructures \cite{Ji09}. 

Subsequently, we investigate the effective transport properties and wave-propagation characteristics of these materials. We demonstrate that the anomalous suppression of volume-fraction fluctuations in hyperuniform two-phase materials endow them with novel and often optimal transport and electromagnetic properties. In the case of phase-inversion-symmetric materials, we show that they indeed achieve a well-known explicit formula for the effective electrical (thermal) conductivity. On the other hand, the stealthy dispersion possesses nearly optimal effective conductivity while being statistically istropic. It is noteworthy that the frequency-dependent effective dielectric constant of any two-phase hyperuniform material cannot have imaginary part, implying that any such material is dissipationless (i.e., transparent) to electromagnetic radiation in the long-wavelength limit. Hence, all of our designed hyperuniform materials possess such characteristics. Moreover, our constructed dispersion is transparent for a range of wavelengths as well. 

It is noteworthy that our tailored composite microstructures can be readily realized by 3D printing and lithographic technologies. We expect that our designs will be potentially useful for energy-saving materials \cite{Hs16}, batteries \cite{Al10} and aerospace applications \cite{Wu13}.

In Sec. 2, we describe the Fourier-space based construction technique to design disordered hyperuniform two-phase materials. In Sec. 3, we employ our construction technique to generate disordered hyperunform two-phase microstructures with prescribed spectral densities. In Sec. 4. we compute the corresponding effective transport properties and wave-propagation characteristics of the designed two-phase materials. In Sec. 5, we offer concluding marks, and discuss potential application and extension of our results.
\section{Fourier-Space Construction Procedure}
\subsection{Algorithm Description}
The microstructure of a random multi-phase material is uniquely determined by the indicator functions $\mathcal{I}^{(i)}({\bf x})$ associated with all of the individual phases defined as
\begin{equation}
\label{eq_5} \mathcal{I}^{(i)}({\bf x}) = \left \{
\begin{array}{c@{\hspace{0.3cm}}c}
1, & {\bf x}~\textnormal{in}~\textnormal{phase}~ i
\\ 0, & \textnormal{otherwise}
\end{array} \right .
\end{equation}
where $i=1,...,q$ and $q$ is the total number of phases \cite{To02}. For statistically homogeneous two-phase materials where there are no preferred centers, the two-point probability function $S_2^{(i)}({\bf r})$ measures the probability of finding two points separated by vector displacement ${\bf r}$ in phase $i$ \cite{To02}. The autocovariance function $\chi_{_{V}}({\bf r})$ is trivially related to $S_2^{(i)}({\bf r})$ via
\begin{equation}
\label{eq_6} \chi_{_{V}}({\bf r}) \equiv S_2^{(i)}({\bf r}) - \phi_i^2,
\end{equation}
where $\phi_i$ is the volume fraction of phase $i$ \cite{To02}. The spectral density $\widetilde{\chi}_{_V}({\bf k})$ is the Fourier transform of the autocovariance function $\chi_{_V}({\bf r})$, where ${\bf k}$ is the wavevector \cite{Za09, Za11a, Za11b, Za11c, Dr15}. In practice we generally deal with finite-sized digitized materials, i.e.,  materials consisting of pixels (square units) in two dimensions or voxels (cubic units) in three dimensions with each pixel (or voxel) entirely occupied by one phase. We apply periodic boundary conditions to materials as approximation of the infinite system that we are interested in.

The Yeong-Torquato stochastic reconstruction procedure \cite{Ye98a, Ye98b} is a popular algorithm to (re)construct digitized multi-phase media from correlation functions in physical (or direct) space. Liu and Shapiro have further employed advanced structure synthesis techniques that utilize a variety of microstructure descriptors in physical space to design functionally graded materials \cite{Li17}. We note that there is a variety of other methods that have been developed to generate or reconstruct microstructures from limited structural information in the direct space; see Refs. \cite{Ro97, Fu08, Ha11, Ta13, Li15, Ca17, Ji07, Ji08, Ta09} and references therein. 

In this paper, we generalize the Yeong-Torquato procedure to construct disordered hyperuniform materials with desirable effective macroscopic properties but from structural information in Fourier (reciprocal) space, i.e., the spectral density $\widetilde{\chi}_{_V}({\bf k})$. Specifically, we define a fictitious ``energy'' $E$ of the system as the squared differences between the target and (re)constructed spectral densities, i.e.,
\begin{equation}
\label{eq_7} E = \sum_{\bf k} [\widetilde{\chi}_{_V}({\bf k})/l^d - \widetilde{\chi}_{_{V,0}}({\bf k})/l^d]^2,
\end{equation}
where the sum is over discrete wave vectors ${\bf k}$, $\widetilde{\chi}_{_{V,0}}({\bf k})$ and $\widetilde{\chi}_{_V}({\bf k})$ are the spectral densities of the target and (re)constructed microstructures, $d$ is the dimension, and $l$ is certain characteristic length of the system used to scale the spectral densities such that they are dimensionless. We employ simulated annealing method \cite{Ye98a} to minimize the energy of the system. We start with random initial configurations with prescribed volume fractions of both phases. At each time step we randomly select one pixel (or voxel) from each of the two phases and attempt to swap them \cite{Ji07, Ji08}. In the later stages of the construction procedure, we apply the different-phase-neighbor-based pixel swap rule, an advanced rule developed previously \cite{Ta09} to improve efficiency of the algorithm and remove random ``noise'' (isolated pixel or small clusters of pixels of the phase of interest) in the media. We update the spectral density of the trial configuration $\widetilde{\chi}_{_V}({\bf k})$ and compute the system energy. We accept the trial pixel swap according to the probability
\begin{equation}
\label{eq_8} p_{acc}(old\rightarrow new) = \textnormal{min}\{1, \textnormal{exp}(-\frac{E_{new}-E_{old}}{T})\},
\end{equation}
where $T$ is the fictitious ''temperature'' of the system that is set initially high and gradually decreases according to a cooling schedule \cite{Ye98a, Ye98b}, and $E_{old}$ and $E_{new}$ are the energies of the system before and after the pixel swap. Trial pixel swaps are repeated and system energy is tracked until it drops below a specified stringent threshold value, which we choose as $10^{-3}$ in this work. 
\subsection{Efficient algorithmic implementation of the construction technique}
In this work, we consider digitized two-phase materials in a square domain in two dimensions subject to periodic boundary conditions and denote the side length of the square domain by $L$. Here we set $L$ to be 300 pixels. For such materials, the wave vector ${\bf k}$ can only take discrete values ${\bf k}=2\pi\times(n_1\widehat{x}+n_2\widehat{y})/L~~(n_1, n_2 \in \mathbb{Z})$, where $\widehat{x}$, $\widehat{y}$ are two orthogonal unit vectors aligned with the boundaries of the square domain. It can be easily shown that the spectral density of such materials can be computed as
\begin{equation}
\label{eq_10} \widetilde{\chi}_{_V}({\bf k})=\frac{1}{A}\widetilde{m}^2({\bf k})|\widetilde{\mathcal{J}}({\bf k})|^2,
\end{equation}
where $A=L^2$ is the area of the system, $\widetilde{m}({\bf k})$ is the Fourier transform of the indicator function $m({\bf r})$ of a pixel and given by
\begin{equation}
\label{eq_11} \widetilde{m}({\bf k}) = \left \{
\begin{array}{c@{\hspace{0.3cm}}c}
\frac{\sin(k_x/2)}{(k_x/2)}\frac{\sin(k_y/2)}{(k_y/2)}, & k_x\neq0,~k_y\neq0
\\ \frac{\sin(k_x/2)}{(k_x/2)}, & k_x\neq0,~k_y=0
\\ \frac{\sin(k_y/2)}{(k_y/2)}, & k_x=0,~k_y\neq0
\\ 1, & k_x=0,~k_y=0
\end{array} \right .,
\end{equation}
and the generalized collective coordinate \cite{To15} $\widetilde{\mathcal{J}}({\bf k})$ is defined as
\begin{equation}
\label{eq_a3} \widetilde{\mathcal{J}}({\bf k})=\sum_{\bf r} [\textnormal{exp}(i{\bf k\cdot r})(\mathcal{I}({\bf r})-\phi)],
\end{equation}
where ${\bf r}$ sums over all the pixel centers, and $\mathcal{I}({\bf r})$ [as defined in Eq. (\ref{eq_5})] and $\phi$ are the indicator function and volume fraction of the phase of interest, respectively. Henceforth, when referring to the properties of the phase of interest, we will drop the subscripts and superscripts for simplicity. In this work we focus on isotropic materials and employ the angular-averaged version $\widetilde{\chi}_{_V}(k)$ of $\widetilde{\chi}_{_V}({\bf k})$ in the energy functional $E$ as defined in Eq. (\ref{eq_7}).

A central issue in the construction procedure is to compute the spectral density of the trial configurations efficiently. Here instead of computing $\widetilde{\chi}_{_V}(k)$ from scratch for every new configuration, we have devised a method that enables one to quickly compute $\widetilde{\chi}_{_V}(k)$ of the new configuration based on the old ones. Specifically we track the generalized collective coordinate $\widetilde{\mathcal{J}}({\bf k})$ at each ${\bf k}$. At the beginning of the simulation, $\widetilde{\mathcal{J}}({\bf k})$ of the initial configuration is explicitly computed. Then for every new trial configuration, the change of $\widetilde{\mathcal{J}}({\bf k})$ only comes from the pixel swap and thus can be updated as follows:
\begin{equation}
\label{eq_13} \widetilde{\mathcal{J}}({\bf k})+d\widetilde{\mathcal{J}}^{new}({\bf k})-d\widetilde{\mathcal{J}}^{old}({\bf k})\rightarrow \widetilde{\mathcal{J}}({\bf k}),
\end{equation}
where
\begin{equation}
\label{eq_14} d\widetilde{\mathcal{J}}^{new}({\bf k})=\textnormal{exp}(i{\bf k\cdot r}_{new}),
\end{equation}
\begin{equation}
\label{eq_15} d\widetilde{\mathcal{J}}^{old}({\bf k})=\textnormal{exp}(i{\bf k\cdot r}_{old}),
\end{equation}
and ${\bf r}_{new}$ and ${\bf r}_{old}$ are the new and old positions of the moved pixel that belongs to the phase of interest. Then $\widetilde{\chi}_{_V}({\bf k})$ of the trial configuration is computed using Eq. (\ref{eq_10}) and subsequently binned according to $k=|{\bf k}|$ in order to obtain the angular-averaged $\widetilde{\chi}_{_V}(k)$. If the trial swap is rejected, $\widetilde{\mathcal{J}}({\bf k})$ of the old configuration can be easily restored by
\begin{equation}
\label{eq_16} \widetilde{\mathcal{J}}({\bf k})-d\widetilde{\mathcal{J}}^{new}({\bf k})+d\widetilde{\mathcal{J}}^{old}({\bf k})\rightarrow \widetilde{\mathcal{J}}({\bf k}).
\end{equation}

Note that the complexity of our algorithm is $O(L^2)$, where $L$ is the linear size of the microstructure. The simulations were performed on an Intel(R) Xeon(R) CPU (E5-2665) with a clock speed of 2.40 GHz, and it took roughly one day to generate a typical microstructure.
\section{Designing disordered hyperuniform two-phase materials with prescribed spectral densities}
\begin{figure}[ht!]
\begin{center}
$\begin{array}{c}\\
\includegraphics[width=0.46\textwidth]{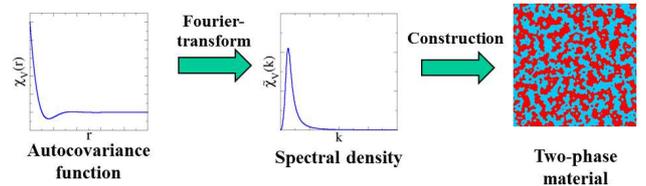}
\end{array}$
\end{center}
\caption{Illustration of the numerical construction procedure to design and generate disordered hyperuniform two-phase materials. The hyperuniformity condition places constraints on $\chi_{_{V}}({\bf r})$ and $\widetilde{\chi}_{_V}({\bf k})$, as given Eq. (\ref{eq_17}).} \label{fig_1}
\end{figure}

Previously certain necessary conditions that autocovariance functions $\chi_{_V}(r)$ have to satisfy so that they can be realized by two-phase materials have been determined \cite{To99, To06}. It is noteworthy that these necessary conditions are not sufficient to guarantee realizablity of $\chi_{_V}(r)$ by two-phase materials, which should be ultimately verified by the successful construction of the targeted spectral densities. Also, certain parameterized autocovariance functions expressible in terms of a set of chosen realizable basis functions have been identified \cite{To99, To06}. Here we utilize this knowledge, but for a completely different purpose, i.e., to design various disordered hyperuniform two-phase materials. Specifically, we first design realizable $\chi_{_{V}}(r)$ with an additional hyperuniform constraint \cite{To16a, To16b}:
\begin{equation}
\label{eq_17} \int_{\mathbb{R}^d}\chi_{_V}({\bf r})d{\bf r}=0,~~\widetilde{\chi}_{_V}(0)=0.
\end{equation}
We then compute the Fourier transform $\widetilde{\chi}_{_{V}}(k)$ of $\chi_{_{V}}(r)$ and employ $\widetilde{\chi}_{_{V}}(k)$ as the target spectral density in the aforementioned Fourier-space construction technique. Subsequently, we carry out this construction technique to construct two-phase materials corresponding to $\widetilde{\chi}_{_{V}}(k)$. The procedure is schematically shown in Fig. \ref{fig_1}.

As a proof-of-concept and for simplicity, we first design a family of disordered hyperuniform materials with phase-inversion symmetry \cite{To02}, i.e., the corresponding microstructures at volume fractions $\phi$ can be generated by inverting the two phases of the microstructures at volume fractions $1-\phi$. Another reason to design these microstructures is that they achieve a well-known explicit formula for effective conductivity, as we show in detail in next section. The scaled autocovariance function $\chi_{_V}(r)/[\phi(1-\phi)]$ of such microstructures is independent of $\phi$ \cite{To06}. Here we explicitly consider constructions at volume fractions $\phi\leq0.5$, and generate microstructures at $\phi>0.5$ by inverting the two phases of the microstructures at $1-\phi$.

\begin{figure}[ht!]
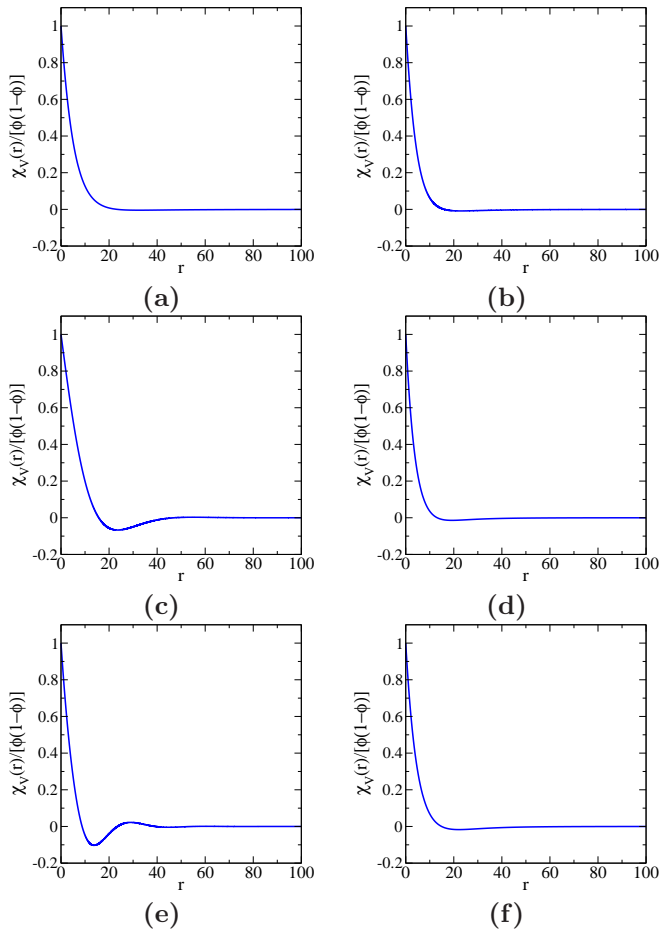

\begin{center}
$\begin{array}{c@{\hspace{0.5cm}}c}\\
\includegraphics[width=0.22\textwidth]{fig2a.eps} &
\includegraphics[width=0.22\textwidth]{fig2b.eps} \\
\mbox{\bf (a)} & \mbox{\bf (b)} \\
\includegraphics[width=0.22\textwidth]{fig2c.eps} &
\includegraphics[width=0.22\textwidth]{fig2d.eps} \\
\mbox{\bf (c)} & \mbox{\bf (d)} \\
\includegraphics[width=0.22\textwidth]{fig2e.eps} &
\includegraphics[width=0.22\textwidth]{fig2f.eps} \\
\mbox{\bf (e)} & \mbox{\bf (f)}\\
\end{array}$
\end{center}
\caption{Realizable autocovariance functions $\chi_{_V}(r)/[\phi(1-\phi)]$ that correspond to hyperuniform two-phase materials, where $\phi$ is the volume fraction of the phase of interest. Functions in (a), (b), (d), (f) are given by Eq. (\ref{eq_21}) with the parameters ($q$, $a$, $b$, $c$) = (5/2, 5, $5\sqrt{15}/2$, 1/4), (3, 4, $4\sqrt{6}$, 1/2), (5, 4, 24, 1/2), and (8, 15, $15\sqrt{14}$, 1/2), respectively. Function in (c) is given by Eq. (\ref{eq_19}) with $q=0.1$ and $a=10.0$. Function in (e) is given by Eq. (\ref{eq_20}) with $q=0.2$, $a=[5(1+\sqrt{3})^{3/2}]/[\sqrt{2}(3+\sqrt{3})]$, and $b=5(\sqrt{2}+\sqrt{6})/2$.} \label{fig_2}
\end{figure}

We start with realizable basis functions for $\chi_{_{V}}(r)$ that were identified previously \cite{To06}. One such example is 
\begin{equation}
\label{eq_19}  \chi_{_V}(r)/[\phi(1-\phi)]=e^{-r/a}\cos(qr),
\end{equation}
where $q$ is the wavenumber associated with the oscillation of $\chi_{_V}(r)$, and $a$ can be considered as the correlation length of the system. In previous work \cite{To16b} it was found that when $qa=1.0$, the corresponding autocovariance function satisfies all the known necessary realizable conditions and the hyperuniformity constraint. Here we set as $q=0.1$ to ensure high enough resolution for the microstructure. Then we set $a=10.0$ such that $qa=1.0$ is satisfied. The resulting $\chi_{_V}(r)$ is shown in Fig. \ref{fig_2}(c).

\begin{figure*}[ht!]
\begin{center}
$\begin{array}{c}\\
\includegraphics[width=0.80\textwidth]{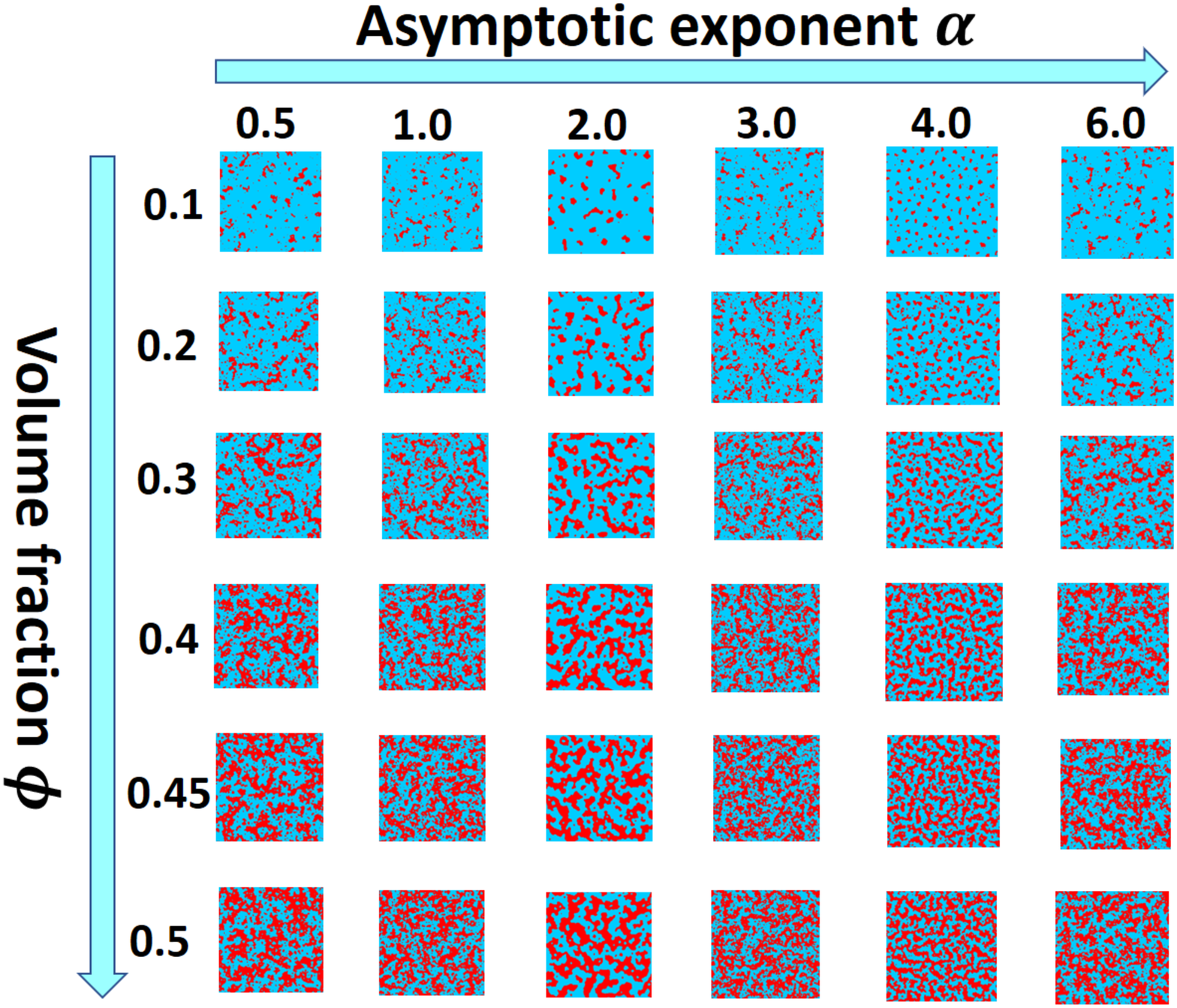} \\
\end{array}$
\end{center}
\caption{Realizations of disordered hyperuniform two-phase materials. From left to right, each column correspdonds to one autocovariance function in Fig. \ref{fig_2}. The quantity $\phi$ is the volume fraction of the phase of interest, and $\alpha$ specifies the asymtotpic behavior of  $\widetilde{\chi}_{_V}(k)$ as $k$ goes to zero, i.e., $\widetilde{\chi}_{_V}(k) \sim k^\alpha$. Note that since these microstructures possess phase-inversion symmetry, the corresponding microstructures at volume fractions $\phi>0.5$ can be generated by inverting the two phases of the microstructures at volume fractions $1-\phi$. Note that depending on the exponent $\alpha$, the volume-fraction variance scaling will behave according to Eq. (\ref{eq_a2}).} \label{fig_3}
\end{figure*}

We compute the corresponding spectral density $\widetilde{\chi}_{_V}(k)$ and employ it to construct two-phase materials. We construct microstructures at volume fractions $\phi = $0.1, 0.2, 0.3, 0.4, 0.45, and 0.5. The constructed microstructures are shown in the third colume of Fig. \ref{fig_3}. Representative target and constructed spectral densities at $\phi = 0.5$ are shown in Fig. \ref{fig_5}(c). It is noteworthy that spectral densities at other values of $\phi$ only differ by certain constants. Note that $\widetilde{\chi}_{_V}(k)$ goes to zero quadratically as $k$ goes to zero, i.e., the scaling exponent $\alpha=2$. In addition, in the opposite asymptotic large-$k$ limit, $\widetilde{\chi}_{_V}(k)$ decays like $1/k^3$, which is consistent with the fact that $\chi_{_V}(r)$ is linear in $r$ for small $r$.

\begin{figure}[ht!]
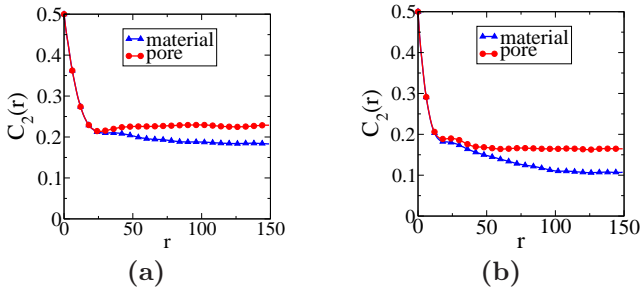

\begin{center}
$\begin{array}{c@{\hspace{1.0cm}}c}\\
\includegraphics[width=0.20\textwidth]{fig4a.eps} &
\includegraphics[width=0.20\textwidth]{fig4b.eps} \\
\mbox{\bf (a)} & \mbox{\bf (b)} \\
\end{array}$
\end{center}
\caption{(a) Two-point cluster function $C_2(r)$ of the microstructure with $\alpha = 2.0$ at $\phi = 0.5$ in Fig. \ref{fig_3}. (b) Two-point cluster function $C_2(r)$ of the microstructure with $\alpha = 4.0$ at $\phi = 0.5$ in Fig. \ref{fig_3}. Note that $C_2(r)$ in (a) decays more slowly than $C_2(r)$ in (b), implying better long-range connectedness of the microstructure with $\alpha = 2.0$ in Fig. \ref{fig_3}.} \label{fig_4}
\end{figure}

\begin{figure}[ht!]
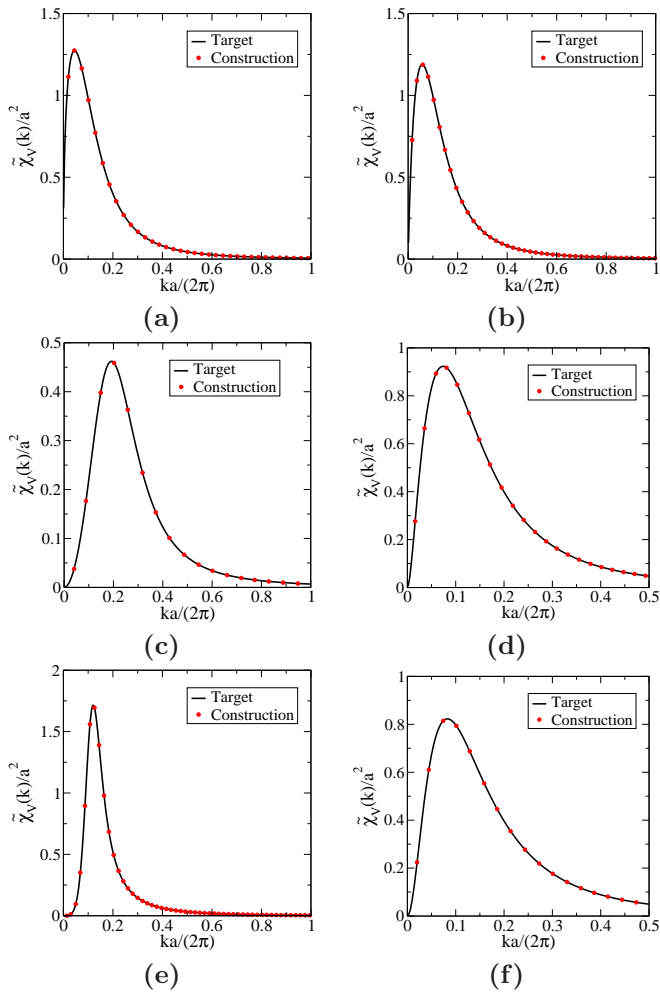

\begin{center}
$\begin{array}{c@{\hspace{0.5cm}}c}\\
\includegraphics[width=0.22\textwidth]{fig5a.eps} &
\includegraphics[width=0.22\textwidth]{fig5b.eps} \\
\mbox{\bf (a)} & \mbox{\bf (b)} \\
\includegraphics[width=0.22\textwidth]{fig5c.eps} &
\includegraphics[width=0.22\textwidth]{fig5d.eps} \\
\mbox{\bf (c)} & \mbox{\bf (d)} \\
\includegraphics[width=0.22\textwidth]{fig5e.eps} &
\includegraphics[width=0.22\textwidth]{fig5f.eps} \\
\mbox{\bf (e)} & \mbox{\bf (f)}\\
\end{array}$
\end{center}
\caption{Representative target and constructed dimensionless spectral densities $\widetilde{\chi}_{_V}(k)/a^2$ that correspond to realizations of hyperuniform microstructures at $\phi = 0.5$ in Fig. \ref{fig_3}, where $a$ is certain characteristic length scale of the systems. It is noteworthy that spectral densities at other values of $\phi$ only differ by certain constants.} \label{fig_5}
\end{figure}

Another basis function investigated previously \cite{To06} is
\begin{equation}
\label{eq_20}  \chi_{_V}(r)/[\phi(1-\phi)]=\frac{1}{2}[e^{-r/a}+e^{-r/b}\cos(qr)],
\end{equation}
where $q$, $a$, and $b$ are parameters (here we choose $a$ as a characteristic length of the system). In previous work \cite{To16b} we find that when $a=[((qb)^2-1)^{1/2}]/[(qb)^2+1]$ and $1<qb\leq(\sqrt{2}+\sqrt{6})/2$, the autocovariance function (\ref{eq_20}) satisfies all the known necessary realizable conditions and the hyperuniformity constraint. Here, similar to the previous case, we choose $q=0.2$. To obtain quartic behavior of $\widetilde{\chi}_{_V}(k)$ near the origin, we set $b=5(\sqrt{2}+\sqrt{6})/2$ such that $qb=(\sqrt{2}+\sqrt{6})/2$ is satisfied, and then set $a=[5(1+\sqrt{3})^{3/2}]/[\sqrt{2}(3+\sqrt{3})]$ to satisfy $a=[((qb)^2-1)^{1/2}]/[(qb)^2+1]$. The resulting $\chi_{_V}(r)$ is shown in Fig. \ref{fig_2}(e). The constructed microstructures are shown in the fifth column of Fig. \ref{fig_3}. Representative target and constructed spectral densities at $\phi = 0.5$ are shown in Fig. \ref{fig_5}(e). It is noteworthy that spectral densities at other values of $\phi$ only differ by certain constants. The constructed spectral density indeed demonstrates hyperuniformity; moreover, $\widetilde{\chi}_{_V}(k)$ is indeed quartic in $k$ around the origin. In addition, in the opposite asymptotic large-$k$ limit, $\widetilde{\chi}_{_V}(k)$ decays like $1/k^3$, which is consistent with the asymptotic behavior of $\chi_{_V}(r)$ for small $r$.

In order to obtain materials with other types of hyperuniformity, i.e., other power laws of $\widetilde{\chi}_{_V}(k)$ around the origin, we introduce a new class of autocovariance function
\begin{equation}
\label{eq_21}  \chi_{_V}(r)/[\phi(1-\phi)]=(c+1)e^{-r/a}-c\frac{b^q}{(r+b)^q},
\end{equation}
where the parameters $q$, $a$, $b$ and $c$ are positive. For this $\chi_{_V}(r)$ to be realizable and hyperuniform, the parameters have to satisfy the following conditions: 
\begin{equation}
\label{eq_a4} (v-2)(v-1)(1+c)a^2=cb^2,
\end{equation}
\begin{equation}
\label{eq_a5} -\frac{1+c}{a}+\frac{cv}{b}<0,
\end{equation}
and
\begin{equation}
\label{eq_a6} \frac{1+c}{a^2}-\frac{cv(1+v)}{b^2}\geq0.
\end{equation}

The relation (\ref{eq_a4}) corresponds to the hyperuniformity constraint, Eq. (\ref{eq_a5}) corresponds to the realizability condition that the first derivative of $\chi_{_V}(r)$ should be negative at $r=0$ \cite{To06}, and Eq. (\ref{eq_a6}) corresponds to the realizability condition that the second derivative of $\chi_{_V}(r)$ should be nonnegative at $r=0$ \cite{To06}. It is noteworthy that this $\chi_{_V}(r)$ scales like $-1/r^q$ for large $r$, which translates into a scaling of $k^{q-d}$ for $\widetilde{\chi}_{_V}(k)$ at small $k$, where $d=2$ is the dimension. Thus by tuning the value of $q$, we can manipulate the type of hyperuniformity that results. Specifically, we aim to obtaining materials with their $\widetilde{\chi}_{_V}(k)$ going to zero as $k$ goes to zero with the following exponents $\alpha$: $1/2$, 1, 3, and, 6, respectively. We find that by setting the parameters ($q$, $a$, $b$, $c$) = (5/2, 5, $5\sqrt{15}/2$, 1/4), (3, 4, $4\sqrt{6}$, 1/2), (5, 4, 24, 1/2), and (8, 15, $15\sqrt{14}$, 1/2), respectively, these desired small-$k$ asymptotic behaviors of $\widetilde{\chi}_{_V}(k)$, i.e., $\alpha=1/2$, 1, 3, and, 6, are achieved. The resulting $\chi_{_V}(r)$ are shown in Fig. \ref{fig_2}(a), (b), (d), and (f), and the constructed microstructures in the first, second, fourth, and sixth columns of Fig. \ref{fig_3}, respectively. Representative target and constructed spectral densities of these microstructures at $\phi = 0.5$ are shown in Fig. \ref{fig_5}(a), (b), (d), and (f), respectively. It is noteworthy that spectral densities at other values of $\phi$ only differ by certain constants. In addition, we note that the microstructures in the first and second columns of Fig. 3 possess the third and second types of volume-fraction variance scaling described in Eq. (\ref{eq_a2}), respectively, and all of the rest microstructures in Fig. 3 possess the first type of scaling described in Eq. (\ref{eq_a2}).

The designed materials shown in Fig. \ref{fig_3} possess a variety of morphologies: as $\phi$ increases, the materials gradually transition from particulate media consisting of isolated ``particles'' to labyrinth-like microstructures. Note that both phases in the microstructures with $\alpha = 2.0$ and $\alpha = 4.0$ at $\phi = 0.5$ in Fig. \ref{fig_3} percolate with nearest-neighbor and next-nearest-neighbor connections (along the pixel edges and corners), which is a singular topological feature for a two-dimensional composite \cite{To02}. Normally, only one phase can percolate (with the other phase being topologically disconnected). It is known that $d$-dimensional ($d\geq2$) two-phase materials that possess phase-inversion symmetry are bicontinuous (i.e., both phases percolate) for $\phi_c<\phi<1-\phi_c$, provided that the percolation threshold $\phi_c<1/2$ \cite{To02}. For example, two-dimensional random checkerboard systems are bicontinuous for $0.4073<\phi<0.5927$ with nearest-neighbor and next-nearest-neighbor connections \cite{To02}. Also, to quantify the differences in long-range topological connectedness of the microstructures with $\alpha = 2.0$ and $\alpha = 4.0$ at $\phi = 0.5$ in Fig. \ref{fig_3}, we have computed their corresponding two-point cluster functions $C_2(r)$, which measure the probability of finding two points separated by $r$ in the same cluster of the phase of interest \cite{To02, Ji09}, as shown in Fig. \ref{fig_4}(a) and (b). A cluster is defined as any topologically connected region of a phase. Clearly the microstructures with $\alpha= 2.0$ is less connected than the one with $\alpha = 4.0$ on large length scales, which is consistent with the observation that the exponentially decaying term in Eq. (\ref{eq_20}) gives rise to clusters of random sizes and shapes \cite{Ji07, Ji08}. 

\begin{figure}[ht!]
\begin{center}
$\begin{array}{c}\\
\includegraphics[width=0.36\textwidth]{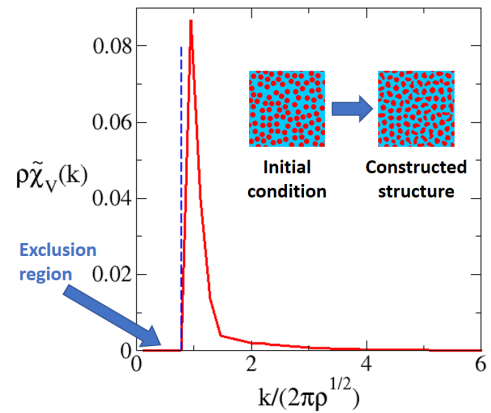}
\end{array}$
\end{center}
\caption{Designed disordered stealthy hyperuniform dispersion that is transparent to long-wavelength electromagnetic radiation and its associated dimensionless spectral density $\rho\widetilde{\chi}_{_{V}}(k)$ (scaled by the number density of the ``particles'' $\rho$). To generate such a pattern, we employ random disk packings as initial conditions and a pixel selection rule that favors the swap of pixels of different phases at the two-phase interphase. We start from very low initial temperature $T_0=10^{-10}$, and only constrain $\widetilde{\chi}_{_{V}}(k)$ to be zero for wavenumbers within the exclusion region, which is shown in the spectral density plot (the region on the left of the blue dash line). Interestingly, this dispersion possesses nearly optimal effective electrical (or thermal) conductivity for a realization with the individual phase conductivity contrast $\sigma_2/\sigma_1 = 10.0$, where $\sigma_1$ and $\sigma_2$ are the electrical (or thermal) conductivities of the ``particle'' and matrix phases, respectively.} \label{fig_6}
\end{figure}

We now consider a construction of hyperuniform materials that does not have phase-inversion symmetry. We employ random disk packings as initial conditions and start from very low initial temperature $T_0=10^{-10}$. We employ a pixel selection rule that favors the swap of pixels of different phases at the two-phase interphase (see Appendix A for detail), and only constrain $\widetilde{\chi}_{_{V}}(k)$ to be zero for wavenumbers within a circular exclusion region around the origin with a radius $K$. We obtain disordered stealthy hyperuniform dispersion at relatively high volume fraction $\phi=0.388$, which appear like the patterns of leopard spots, as shown in Fig. \ref{fig_6}. Here $K$ is chosen such that $K/(2\pi\rho^{1/2})\leq0.864$, where $\rho$ is the number density of the ``particles'' (We note that for a microstructure with a linear size of 300 pixels, the distance between neighboring $k$ points in the Fourier space is $2\pi/300$, and if we choose a bin size roughly twice as large as this distance for $k$ to compute $\widetilde{\chi}_{_{V}}(k)$, the exclusion region with a radius 0.864 includes $k$ points within the first 5 bins). This example serves to demonstrate that by varying the initial conditions and cooling schedule,  there is a wide diversity of microstructures that can be generated by our construction technique. Note that this dispersion is transparent to electromagnetic radiation with wavenumbers smaller than $K$ in the single-scattering regime. Such materials should also be transparent for a range of wavelengths in the multiple-scattering regime when the incident wavenumber of the radiation is less than about $K/4$ \cite{Le16}.
\section{Transport and wave-propagation properties of the designed materials}
\subsection{Effective conductivity of the designed materials}
In this section, we first compute the effective electric (or thermal) conductivity $\sigma_e$ of the aforementioned designed materials. According to the homogenization theory \cite{To02}, the effective conductivity $\sigma_e$ is defined through the averaged Ohm's (or Fourier's) law: 
\begin{equation} 
\label{eq_a7} \langle\rvector{J}({\bf x})\rangle=\sigma_e\langle\rvector{E}({\bf x})\rangle,
\end{equation}
where angular brackets denote an ensemble average, $\langle\rvector{J}({\bf x})\rangle$ is the average flux and $\langle\rvector{E}({\bf x})\rangle$ is the average field.

We consider the case where the phase conductivity contrast of the two individual phases $\sigma_2/\sigma_1$ is 10.0. To compute the effective conductivities $\sigma_e$ of the constructed digitized materials, we employ the first-passage-time simulation techniques \cite{To99b, Ge08, Zh16}. Specifically we release over $10^5$ random walkers to sample each material and record the mean squared displacement $\langle R^2(t)\rangle$ of the random walkers at sufficiently large $t$. Then $\sigma_e$ is computed as
\begin{equation}
\label{eq_26} \sigma_e=\lim_{t\rightarrow\infty}\langle R^2(t)\rangle/(2dt),
\end{equation}
where $d$ is the dimension. 

\begin{figure}[ht!]
\begin{center}
$\begin{array}{c}\\
\includegraphics[width=0.42\textwidth]{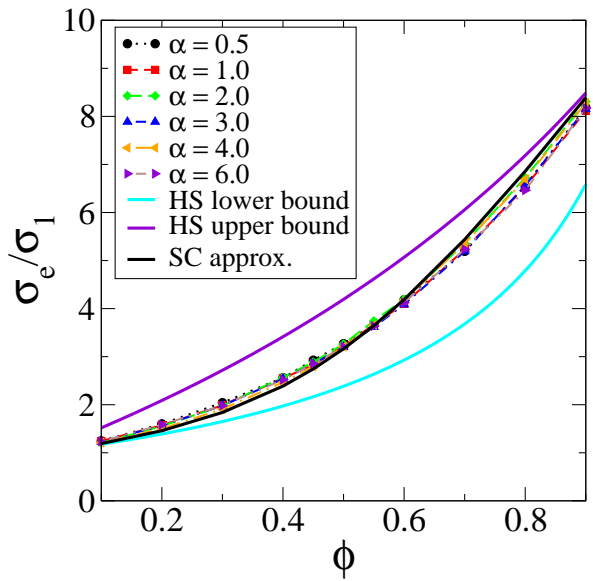} \\
\end{array}$
\end{center}
\caption{The effective conductivities of the designed microstructures across different values of $\alpha$ and $\phi$ in Fig. \ref{fig_3}, as computed from the first-passage-simulation techniques. Here we consider the case where the contrast of phase conductivities $\sigma_2/\sigma_1$ is 10.0, and $\phi$ is the volume fraction of phase 2, which is targeted in the construction technique. The HS two-point bounds on $\sigma_e$ as well as the SC approximation as described by Eq. (\ref{eq_28}) are also plotted alongside the simulation results. The effective conductivities $\sigma_e$ of these microstructures indeed agree well with the SC approximation.} \label{fig_7}
\end{figure}

Torquato \cite{To02} has derived a ``strong-contrast'' expansion of $\sigma_e$ that perturbs around the microstructures that realize the well-known self-consistent (SC) approximation for effective conductivity:
\begin{equation}
\label{eq_27} 
\begin{split}
&\phi_2\frac{\sigma_e+(d-1)\sigma_1}{\sigma_e-\sigma_1}+\phi_1\frac{\sigma_e+(d-1)\sigma_2}{\sigma_e-\sigma_2}\\&=2-d-\sum_{n=3}^{\infty}[\frac{A_n^{(2)}}{\phi_2}\beta_{21}^{n-2}+\frac{A_n^{(1)}}{\phi_1}\beta_{12}^{n-2}],
\end{split}
\end{equation}
where $A_n^{(p)}$ is the $n$-point parameter, and $\beta_{12}=-\beta_{21}=(\sigma_1-\sigma_2)/(\sigma_1+\sigma_2)$. If we truncate Eq. (\ref{eq_27}) after third-order terms and set $A_3^{(2)}=\phi_1\phi_2^2$, Eq. (\ref{eq_27}) reduces to the SC approximation, which in two dimensions is given by
\begin{equation}
\label{eq_28} \phi_2\frac{\sigma_e+\sigma_1}{\sigma_e-\sigma_1}+\phi_1\frac{\sigma_e+\sigma_2}{\sigma_e-\sigma_2}=0.
\end{equation}
The reader is referred to Ref. \cite{To02} for more details about this SC approximation. Milton showed that multiscale hierarchical self-similar microstructures realize the SC approximation \cite{Mi85}. Torquato and Hyun further found a class of periodic, single-scale dispersions that realize this approximation \cite{To01}. Here we discover that a family of disordered single-scale microstructures can also realize this approximation, as shown in Fig. \ref{fig_3}. The effective conductivity results for these microstructures across phase volume fractions $\phi$ are shown in Fig. \ref{fig_7}. The Hashin-Shtrikman (HS) two-point bounds on $\sigma_e$ as well as the SC approximation as described by Eq. (\ref{eq_28}) \cite{To02} are also plotted alongside the simulation results. The effective conductivities $\sigma_e$ of these microstructures indeed agree well with the SC approximation. This agreement demonstrates our ability to construct microstructures with targeted transport properties. 

Now we determine the dimensionless effective conductivity $\sigma_e/\sigma_1$ of the disordered stealthy hyperuniform dispersion described in Fig. \ref{fig_6}. Again, using first-passage time techniques \cite{To99b, Ge08, Zh16}, we find that $\sigma_e/\sigma_1=4.92$. The corresponding HS upper and lower bounds on the dimensionless effective conductivity for any two-phase material with such a phase volume fraction and phase conductivities are determined to be 5.18 and 3.01, respectively. Thus, we see that the effective conductivity $\sigma_e$ of the stealthy dispersion is close to the upper bound, which means that this disordered stealthy dispersion possesses a nearly optimal effective conductivity.

\subsection{Frequency-dependent effective dielectric constant of the designed materials}
Here we consider the determination of the frequency-dependent effective dielectric constant $\varepsilon_e(k_1)$ of the constructed microstructures, which we treat as two-phase dielectric random media with \textit{real} phase dielctric constants $\varepsilon_1$ and $\varepsilon_2$. Here $k_1$ is the wavenumber of the wave propagation through phase $1$. In this case, the attenuation of the waves propagating through the effective medium is due purely to scattering, not absorption \cite{Re08}. We assume that the wavelength of the propagation wave is much larger than the scale of inhomogeneities in the medium. We are interested in the effective dielectric constant $\varepsilon_e(k_1)$ associated with the \textit{homogenized} dynamic dielectric problem \cite{To02, Re08}. Rechtsman and Torquato \cite{Re08} have derived a two-point approximation based on the strong-contrast expansion to estimate $\varepsilon_e$ for two- and three-dimensional microstructures with a percolating phase 2 and $\varepsilon_2\geq\varepsilon_1$:
\begin{equation}
\label{eq_29} \frac{\varepsilon_1-\varepsilon_2}{\varepsilon_1+\varepsilon_2}\phi_1^2[\frac{\varepsilon_e-\varepsilon_2}{\varepsilon_e+\varepsilon_2}]^{-1}=\phi_1-A_2^{(1)}\times[\frac{\varepsilon_1-\varepsilon_2}{\varepsilon_1+\varepsilon_2}],
\end{equation}
where $A_2^{(1)}$ is the two-point parameter that is an integral over the autocovariance function weighted with gradients of the relevant Green's functions. The latter was explicitly represented in three dimensions, but not in two dimensions. It can be shown (see Appendix B for details) that $A_2^{(1)}$ in two dimensions is given by
\begin{equation}
\begin{split}
& A_2^{(1)}=[-\gamma k_1^2\int_{0}^{\infty}\chi_{_V}(r)rdr-k_1^2\ln k_1\int_{0}^{\infty}\chi_{_V}(r)rdr
\\& -k_1^2\int_{0}^{\infty}\chi_{_V}(r)r\ln(r/2)dr]+i\frac{\pi}{2}k_1^2\int_{0}^{\infty}\chi_{_V}(r)rdr+O(k_1^4\ln k_1),
\end{split}
\label{eq_30}
\end{equation}
where $\gamma\approx0.577216$ is the Euler's constant.

\begin{figure}[ht!]
\begin{center}
$\begin{array}{c}\\
\includegraphics[width=0.42\textwidth]{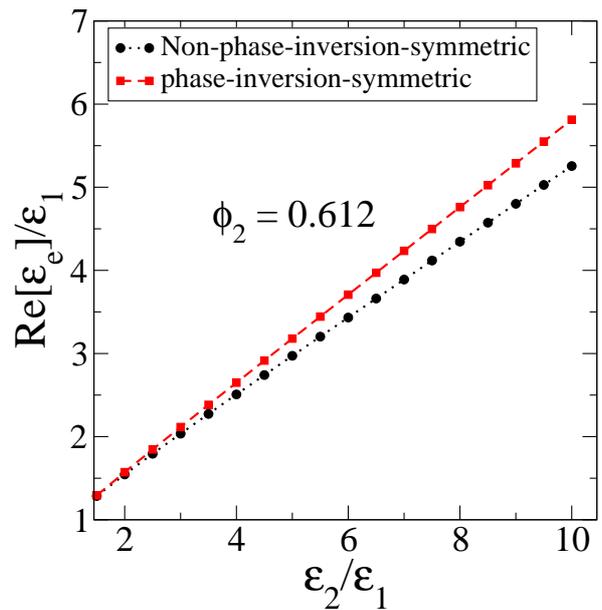} \\
\end{array}$
\end{center}
\caption{Real part of the effective dielectric constant Re$[\varepsilon_e]$ of two hyperuniform materials: a phase-inversion-symmetric case of a realization of Eq. (\ref{eq_19}) and a non-phase-inversion-symmetric one in Fig. \ref{fig_6} as a function of dielectric-contrast ratio $\varepsilon_2/\varepsilon_1$ at volume fraction $\phi_2=0.612$ and wave number $k_1=2\pi/(10a)$. In order to calculate $A_2^{(1)}$ for each microstructure, Eq. (\ref{eq_30}) is used. Clearly Re$[\varepsilon_e]$ differ for these two systems across different values of $\varepsilon_2/\varepsilon_1$.} \label{fig_8}
\end{figure}

The imaginary part of $\varepsilon_e(k_1)$ accounts for attenuation (losses) due to incoherent multiple scattering in a typical disordered two-phase material \cite{Re08}. Because of the sum rule Eq. (\ref{eq_17}) on any $\chi_{_V}(r)$ that corresponds to disordered hyperuniform two-phase materials, it immediately follows that the imaginary part of $A_2^{(1)}$ in Eq. (\ref{eq_30}), and hence $\varepsilon_e(k_1)$ in Eq. (\ref{eq_29}) vanish in the long-wavelength limit for any such materials. As a result, any hyperuniform material is transparent to electromagnetic radiation, i.e., dissipationless in the long-wavelength limit according to the approximation (\ref{eq_30}). This is because the attenuation of propagating waves in such composite materials with real phase dielectric constants can only come from scattering, as mentioned above. Also, the first two terms of the real part of $A_2^{(1)}$ in Eq. (\ref{eq_30}) vanish for such materials, while the remaining lowest-order term $-k_1^2\int_{0}^{\infty}\chi_{_V}(r)r\ln(r/2)dr$ generally does not vanish and depend on the exact form of $\chi_{_V}(r)$. We compute the real part of $\varepsilon_e$ for two hyperuniform materials: a phase-inversion-symmetric case of a realization of Eq. (\ref{eq_19}) and a non-phase-inversion-symmetric one in Fig. \ref{fig_6} at $\phi_2=0.612$, where phase 2 is the percolating matrix phase. We take $k_1=2\pi/(10a)$, where $a=10.0$ is the characteristic length in Eq. (\ref{eq_19}). We first compute $A_2^{(1)}$ from Eq. (\ref{eq_30}), and then compute $\varepsilon_e$ from Eq. (\ref{eq_29}) using $A_2^{(1)}$. The results are shown in Fig. \ref{fig_8}. Clearly the real part of $\varepsilon_e$ differ for these two systems across values of $\varepsilon_2/\varepsilon_1$. At last, we note that the frequency-dependent dielectric constant problem is demonstrated to be equivalent to the static effective conductivity problems as the wavenumber of the propagation wave goes to zero \cite{To02}.
 
\section{Conclusions and discussion}
In this work, we developed for the first time a Fourier-space numerical construction procedure to design at will a wide class of disordered hyperuniform two-phase materials. These materials possess anomalous suppression of volume-fraction fluctuations at large length scales, which endow them with novel and often optimal transport and electromagnetic properties as we demonstrated. Our designed phase-inversion-symmetric materials possess various morphologies and different levels of topological connectedness, as revealed by the two-point cluster function. Moreover, they indeed achieve a well-known explicit formula for the effective electrical (thermal) conductivity. On the other hand, our designed disordered stealthy hyperuniform dispersion possesses nearly optimal effective conductivity, while being fully isotropic. Such materials can sometimes offer advantages over periodic structures with high crystallographic symmetries where the physical properties can be anisotropic, such as has been shown in the case of photonic materials \cite{Fl09, Ma13}. All of our designed hyperuniform materials are dissipationless (i.e., transparent to electromagnetic radiation) in the long-wavelength limit, which is a common characteristic of hyperuniform materials. Moreover, our dispersion is also transparent to electromagnetic radiation for a range of wavelengths.

In the present paper, we focused on the design of two-dimensional hyperuniform two-phase materials with prescribed spectral densities, but it is noteworthy that with slight modification our Fourier-space numerical construction technique can be readily applied in three dimensions to design disordered hyperuniform microstructures, which are expected to be distinctly different from their two-dimesional counterparts. For example, bicontinuous microstructures are much more common in three dimensions \cite{To02}. Moreover, all of our in-silico designed microstructures can be readily realized by 3D printing and lithographic technologies \cite{Va13}. Also, in principle there is no constraint on the types of constituent materials that can be used in these composite materials as long as the constituents (phases) are arranged in a hyperuniform fashion. In addition, we note that a two-phase material can be viewed as a special case of random scalar fields, and our results can be further generalized to design hyperuniform scalar fields, which has received recent attention \cite{To16a, Ma17}.

Since all of our hyperuniform material designs are dissipationless in the long-wavelength limit, they will be potentially useful for energy-saving materials that prevent heat accumulation by allowing the free transmission of infrared radiation \cite{Hs16}. In addition, by employing a phase-change material as the particle phase and graphite as the matrix phase in our designed disordered stealthy hyperuniform dispersion, one could fabricate phase-change composites with high thermal conductivity \cite{Pi08}. Such composite materials can absorb and distribute heat efficiently, which is crucial for the normal operation of battery packs \cite{Al10} and spacecrafts \cite{Wu13}.

A natural extension of this work will be the statistical characterization of our designed disordered hyperuniform two-phase systems by computing a host of different types of statistical correlation functions. This not only includes various types of two-point correlation functions, e.g., pore-size functions, lineal-path functions, surface-surface correlation functions, to name a few, but their higher-order (three-point) generalizations as well \cite{To02, Gi15}. Moreover, to gain further insight into the potential of these constructed microstructures, it will be beneficial to carry out a comprehensive study to estimate other transport, thermal, mechanical, photonic and phononic properties as well as effective reaction rates of these composites. 

Moreover, we note that identifying and utilizing process-structure-property relationships to design and manufacture novel materials with desirable properties is a holy grail of materials science. The emergence of Integrated Computational Materials Engineering (ICMG) has greatly accelerated this process by integrating materials science and automated design \cite{Ca17}. Our present results demonstrate that by designing hyperuniform microstructures with tunable spectral densities, which are then automatically generated at the mesoscale, we can control the effective physical properties of the materials. By combining our construction technique with existing material models and data infrastructures \cite{Wo16}, one may be able to create new powerful ICMG platforms to efficiently design optimized materials for various applications. 
\section*{Acknowledgments}
The authors are very grateful to Dr. Ge Zhang for his careful reading of the manuscript and Dr. Yang Jiao for his helpful discussion. This work was supported by the National Science Foundation under Award No. CBET-1701843.

\appendix
\section{Pixel-selection rule for construction of disordered stealthy hyperuniform dispersion}
To construct disordered stealthy hyperuniform dispersion shown in Fig. \ref{fig_6}, we modify the different-phase-neighbor-based (DPN-based) pixel selection rule proposed in Ref. \cite{Ta09}. In two dimensions, each pixel has 8 neighbors, and we divide the pixels of each phase into different sets $S_i$ based on the number of neighboring pixels $i$ in a different phase that they have. For example, if we consider a two-phase medium consisting of blue and red pixels, we divide the blue pixels into different sets based on the number of neighboring red pixels that they have. 

In each pixel-swap iteration, for each phase in the medium, a set $S_i$ is first selected according to $p(S_i)$, which is given by 
\begin{equation}
\label{eq_s1} p(S_i) = \left \{
\begin{array}{l@{\hspace{0.3cm}}c}
0.6, & i=M,
\\ wA(S_i)(i+1)^4, & 0\leq i<M,
\end{array} \right. 
\end{equation} 
where $M$ is the maximum number of different-phase neighbors of a pixel in the phase of interest, $A(S_i)$ is the number of pixels in the phase of interest with $i$ different-phase neighbors, and $w$ is the normalization factor given by
\begin{equation}
\label{eq_s2} w = (1-0.6)/[\sum_{i=0}^{M-1} A(S_i)(i+1)^4].
\end{equation} 
Then for each phase a pixel is randomly selected from the corresponding chosen $S_i$, and the two selected pixels belonging to the two different phases are swapped, generating a new trial microstructure.

\section{Derivation of the two-point parameter associated with the frequency-dependent dielectric constant in two dimensions}
Here we apply the general formalism derived in Ref. \cite{Re08} to two dimensions. Specifically, the dyadic Green's function is given by 
\begin{equation}
\label{eq_s3} \rttensor{G}(\rvector{r},\rvector{r'}) = -\frac{\rttensor{I}}{2\sigma_1}\delta(\rvector{r}-\rvector{r'})+\rttensor{G}_1(\rvector{r},\rvector{r'})\rttensor{I}+\rttensor{G}_2(\rvector{r},\rvector{r'})\hat{\rvector{r}}\hat{\rvector{r}}.
\end{equation}
where $\hat{\rvector{r}}$ is a unit vector directed from $\rvector{r'}$ towards $\rvector{r}$, $\rttensor{I}$ is the unit tensor, $\delta(\rvector{r}-\rvector{r'})$ is the Dirac delta function, $\sigma_1=k_1^2$ ($k_1$ is the wavenumber of the wave propagating through phase 1), and
\begin{equation}
\label{eq_s4} \rttensor{G}_1(\rvector{r},\rvector{r'})=\frac{i}{4}[H_0^{(1)}(k_1r)-\frac{H_1^{(1)}(k_1r)}{k_1r}],
\end{equation}
\begin{equation}
\label{eq_s5} \rttensor{G}_2(\rvector{r},\rvector{r'})=\frac{i}{4}[\frac{H_1^{(1)}(k_1r)}{k_1r}+\frac{1}{2}H_2^{(1)}(k_1r)-\frac{1}{2}H_0^{(1)}(k_1r)].
\end{equation}
Here $H_i^{(1)}(k_1r)$ is the $i$-th order Hankel function of the first kind. Note that the Green's function $\rttensor{G}(\rvector{r},\rvector{r'})$ solves the following partial differential equation:
\begin{equation}
\label{eq_s6} \nabla\times\nabla\times\rttensor{G}(\rvector{r},\rvector{r'})-\sigma_1\rttensor{G}(\rvector{r},\rvector{r'}) = \rttensor{I}\delta(\rvector{r}-\rvector{r'}).
\end{equation}
The two-point parameter $A_2^{(1)}$ associated with the frequency-dependent dielectric constant $\varepsilon_e(k_1)$ is given by
\begin{equation}
A_2^{(1)}=k_1^2\int\textnormal{Tr}[\rttensor{H}(\rvector{r})]\chi_{_V}(\rvector{r})d\rvector{r},
\label{eq_s7}
\end{equation}
where $\chi_{_V}(\rvector{r})$ is the autocovariance function, Tr$[\rttensor{H}(\rvector{r})]$ is the trace of $\rttensor{H}(\rvector{r})$, and $\rttensor{H}(\rvector{r})$ is the principle value of the Green's function given by
\begin{equation}
\rttensor{H}(\rvector{r},\rvector{r'}) = \rttensor{G}_1(\rvector{r},\rvector{r'})\rttensor{I}+\rttensor{G}_2(\rvector{r},\rvector{r'})\hat{\rvector{r}}\hat{\rvector{r}}.
\label{eq_s8}
\end{equation}
Substituting Eqs. (\ref{eq_s4}) and (\ref{eq_s5}) into Eq. (\ref{eq_s8}), we obtain the explicit expression for $A_2^{(1)}$, which is Eq. (\ref{eq_30}).

\section*{References}

\end{document}